\documentclass[a4paper]{article}

\usepackage{icrc2013}

\usepackage{upgreek}

\usepackage{amsmath}
\usepackage[english]{babel}

\title{The energy-spectrum of light primaries in the range from 10$\mathbf{^{16.6}}$ to 10$\mathbf{^{18.2} \, \mathrm{\textbf{eV}}}$}

\newcommand{\etal}{\MakeLowercase{\textit{et al. }}} 
\shorttitle{Sven Schoo \etal Energy-spectrum of light primaries}

\authors{
S.~Schoo$^{1}$,
W.D.~Apel$^{1}$,
J.C.~Arteaga-Vel\'azquez$^{2}$,
K.~Bekk$^{1}$,
M.~Bertaina$^{3}$,
J.~Bl\"umer$^{1,4}$,
H.~Bozdog$^{1}$,
I.M.~Brancus$^{5}$,
E.~Cantoni$^{3,6,a}$,
A.~Chiavassa$^{3}$,
F.~Cossavella$^{4,b}$,
C.~Curcio$^{3}$,
K.~Daumiller$^{1}$,
V.~de~Souza$^{7}$,
F.~Di~Pierro$^{3}$,
P.~Doll$^{1}$,
R.~Engel$^{1}$,
J.~Engler$^{1}$,
B.~Fuchs$^{4}$,
D.~Fuhrmann$^{8,c}$,
H.J.~Gils$^{1}$,
R.~Glasstetter$^{8}$,
C.~Grupen$^{9}$,
A.~Haungs$^{1}$,
D.~Heck$^{1}$,
J.R.~H\"orandel$^{10}$,
D.~Huber$^{4}$,
T.~Huege$^{1}$,
\mbox{K.-H.}~Kampert$^{8}$,
D.~Kang$^{4}$, 
H.O.~Klages$^{1}$,
K.~Link$^{4}$, 
P.~{\L}uczak$^{11}$,
M.~Ludwig$^{4}$,
H.J.~Mathes$^{1}$,
H.J.~Mayer$^{1}$,
M.~Melissas$^{4}$,
J.~Milke$^{1}$,
B.~Mitrica$^{5}$,
C.~Morello$^{6}$,
J.~Oehlschl\"ager$^{1}$,
S.~Ostapchenko$^{1,d}$,
N.~Palmieri$^{4}$,
M.~Petcu$^{5}$,
T.~Pierog$^{1}$,
H.~Rebel$^{1}$,
M.~Roth$^{1}$,
H.~Schieler$^{1}$,
F.G.~Schr\"oder$^{1}$,
O.~Sima$^{12}$,
G.~Toma$^{5}$,
G.C.~Trinchero$^{6}$,
H.~Ulrich$^{1}$,
A.~Weindl$^{1}$,
J.~Wochele$^{1}$,
J.~Zabierowski$^{11}$, KASCADE-Grande Collaboration
}

\afiliations{
$^1$ Institut f\"ur Kernphysik, KIT - Karlsruher Institut f\"ur Technologie, Germany\\
$^2$ Universidad Michoacana, Instituto de F\'{\i}sica y Matem\'aticas, Morelia, Mexico\\
$^3$ Dipartimento di Fisica, Universit\`a degli Studi di Torino, Italy\\
$^4$ Institut f\"ur Experimentelle Kernphysik, KIT - Karlsruher Institut f\"ur Technologie, Germany\\
$^5$ Horia Hulubei National Institute of Physics and Nuclear Engineering, Bucharest, Romania\\
$^6$ Osservatorio Astrofisico di Torino, INAF Torino, Italy\\
$^7$ Universidade S$\tilde{a}$o Paulo, Instituto de F\'{\i}sica de S\~ao Carlos, Brasil\\
$^8$ Fachbereich Physik, Universit\"at Wuppertal, Germany\\
$^9$ Department of Physics, Siegen University, Germany\\
$^{10}$ Dept. of Astrophysics, Radboud University Nijmegen, The Netherlands\\
$^{11}$ National Centre for Nuclear Research, Department of Cosmic Ray Physics, Lodz, Poland\\
$^{12}$ Department of Physics, University of Bucharest, Bucharest, Romania\\
\scriptsize{
$^{a}$ now at: Istituto Nazionale di Ricerca Metrologia, INRIM, Torino; \\
$^{b}$ now at: Max-Planck-Institut Physik, M\"unchen, Germany; \\
$^{c}$ now at: University of Duisburg-Essen, Duisburg, Germany;  \\
$^{d}$ now at: University of Trondheim, Norway.
}
}
\email{sven.schoo@kit.edu}

\abstract{Data of the Grande extension of the KASCADE experiment allows us to study extensive air showers induced by primary cosmic rays with energies above 10$^{16} \, \mathrm{eV}$. The energy of an event is estimated in terms of the number of charged particles (N$_{\mathrm{ch}}$) and the number of muons (N$_{\mathrm{\upmu}}$) measured at an altitude of ~110$\, \mathrm{m} \, \mathrm{a.s.l.}$ While a combination of the two numbers is used for the energy, the ratio defines the primary mass (group).
The spectrum of the combined light and medium mass components, recently measured with KASCADE-Grande, was found
to be compatible with both a single power-law and a broken power-law in the energy range between 10$^{16.3}$ and 10$^{18} \, \mathrm{eV}$.
In this contribution we will present the investigation of possible structures in the spectrum of light primaries with
increased statistics both from a larger data set including more recent measurements
and by using a larger fiducial area than in the previous study. With the better statistical accuracy and with optimized selection criteria for enhancing light primaries we have found evidence for a hardening (ankle) of the spectrum of the light component at an energy of $10^{17.08 \pm 0.08} \, \mathrm{eV}$.
}
\keywords{ KASCADE-Grande, ultra-high energy cosmic rays, air-showers }

\begin{document}
\maketitle
\vspace*{-1.4cm}
\section{Introduction}
There are two major features in the spectrum of cosmic rays: the \emph{knee} at an energy of around $4 \times 10^{15} \, \mathrm{eV}$; and the \emph{ankle} at about $4 \times 10^{18} \, \mathrm{eV}$. It is commonly believed that cosmic rays with energies below the knee originate from sources within our galaxy, whereas cosmic rays with energies above the ankle are believed to be of extragalactic origin. 
In these models, the transition from galactic to extragalactic origin of cosmic rays is generally expected (see e.g.~\cite{Blasi12,Aloisio12,Wolfendale}) to take place in the energy range between the heavy knee and the ankle, i.e. between $10^{17}$ and 10$^{19} \, \mathrm{eV}$~\cite{bergman07}. Most of the models assume that the extragalactic component is dominated by light primaries. Therefore, a contribution of extragalactic cosmic rays should be visible as a hardening of the energy spectrum of light elements. The KASCADE-Grande experiment~\cite{Apel2010202} allows us to study air-showers induced by cosmic rays having a primary energy between 10$^{16}$ and 10$^{18} \, \mathrm{eV}$. The data recorded with KASCADE-Grande can be used to estimate both the primary energy and the mass of an incident particle on an event-by-event basis using the reconstructed total number of charged particles ($N_{\mathrm{ch}}$) and the total number of muons ($N_{\mathrm{\upmu}}$) reaching detector level at a mean atmospheric depth of 1022$\mathrm{g/cm}^{2}$. Covering an energy range being of high interest for the transition, the present study aims to search for structures in the spectrum of light primaries using the reconstruction and analysis procedures described below and in more detail in~\cite{prl107, Apel2012183}.
\vspace*{-0.4cm}
\section{Experimental setup}
\begin{figure}
\includegraphics[width=\columnwidth]{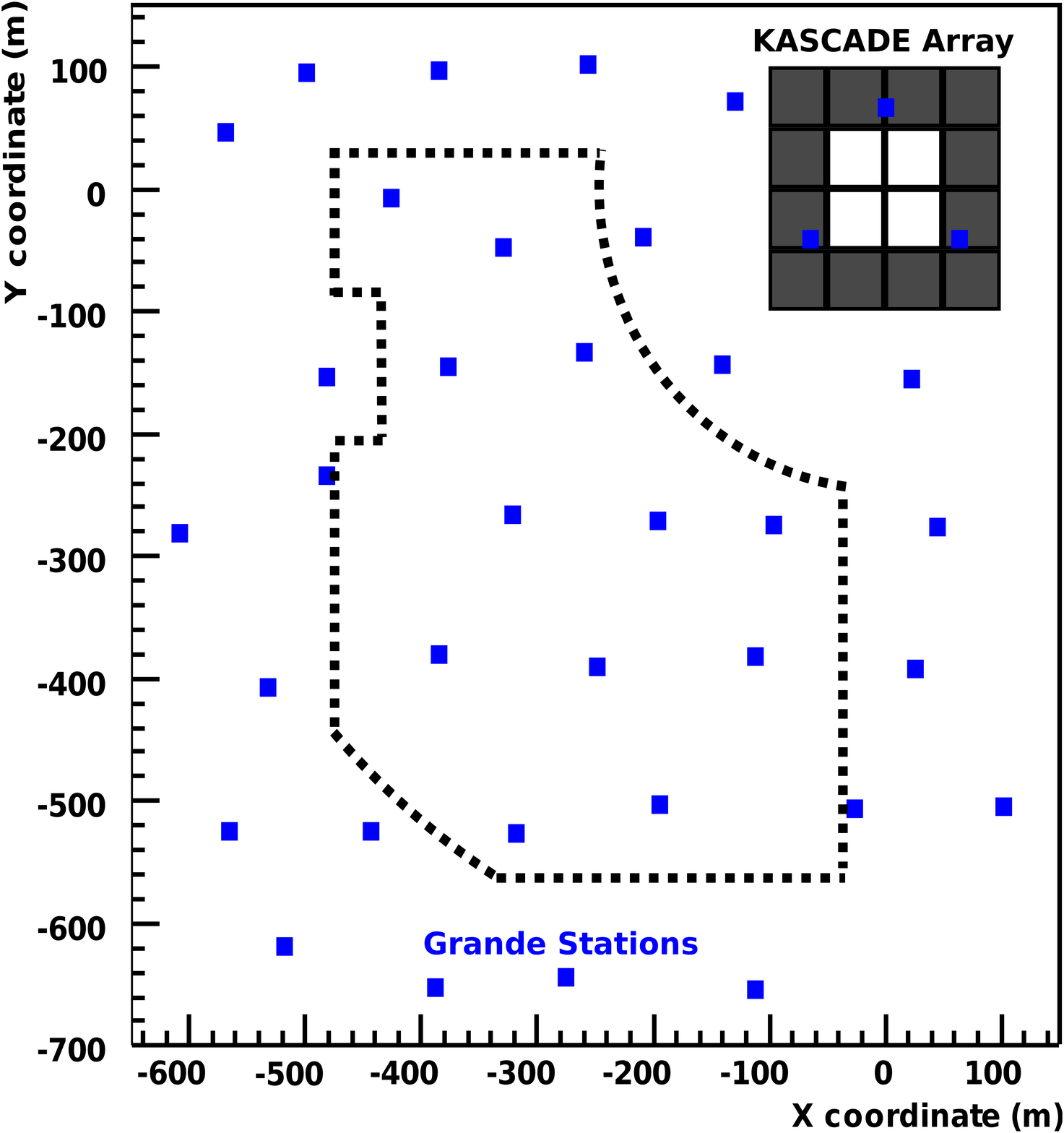}
\caption{Layout of the KASCADE-Grande experiment. The area used in this analysis is indicated by the dotted line. The Grande stations (rectangles) had to be arranged to fit between the buildings. Therefore the stations are irregularly distributed and the missing station at around $(-600, \ -100) \, \mathrm{m}$ is the reason for the rectangular cut between $-80 \, \mathrm{m} \leq Y \leq -200 \, \mathrm{m}$. The original KASCADE array is located in the upper right corner.
}
\label{figLayoutKG}
\end{figure}

The layout of the KASCADE-Grande experiment is displayed in Fig.~\ref{figLayoutKG}.
The Grande array consists of 37 scintillation detector stations distributed over an area of $700 \times 700 \, \mathrm{m}^2$. Each detector has a sensitive area of $10 \, \mathrm{m}^2$. The energy threshold for a charged particle to be registered is about $3 \, \mathrm{MeV}$.
The arrival direction and the core position are first estimated using the arrival times and particle densities measured by the KASCADE-Grande stations. The number of charged particles at observation level is then reconstructed together with the final value of the core position by a fit of an appropriate lateral
distribution function to the particle densities in the KASCADE-Grande detectors. The final value of the arrival direction is obtained by a time fit using the final
core position (See~\cite{Apel2010202} for details).

Equipped with 192 shielded scintillation detectors ($3.2 \, \mathrm{m}^2$ each), the original KASCADE array is used to derive the total number of muons in a similar way using the core position supplied by KASCADE-Grande. The needed energy for vertical incident muons to pass the iron (4$\, \mathrm{cm}$) and lead (10$\, \mathrm{cm}$) shielding is about $230 \, \mathrm{MeV}$.

\begin{figure}
\includegraphics[width=\columnwidth]{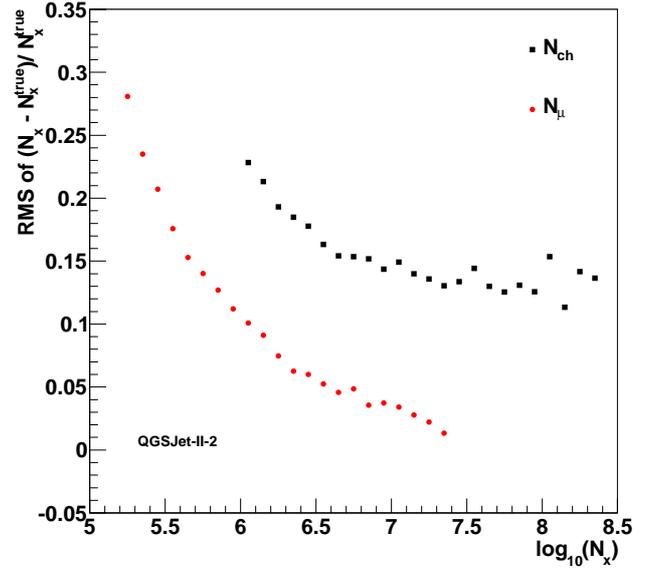}
\caption{Reconstruction accuracies for $N_{\mathrm{ch}}$ and $N_{\mathrm{\upmu}}$. Only events with core positions inside the area shown in Fig.~\ref{figLayoutKG} and zenith angles below $40^{\circ}$ are taken into account.} 
\label{figNchNmuAccur}
\end{figure}

The reconstruction accuracies for $N_{\mathrm{ch}}$ and $N_{\mathrm{\upmu}}$ are shown in Fig.~\ref{figNchNmuAccur}. They are calculated using Monte Carlo simulations. For the simulations, the CORSIKA code~\cite{CORSIKA} was used employing the QGSJET-II-2 high energy hadronic interaction model~\cite{qgsjet06} in case of high energy interactions and Fluka (version 2002.4)~\cite{fluka} in case of interactions at low energies. Using only events with core positions inside the area shown in Fig.~\ref{figLayoutKG} and zenith angles below $40^{\circ}$, the reconstruction accuracy for $N_{\mathrm{ch}} \approx 10^{6}$ is about $22 \, \%$. Above $10^{6.6}$ charged particles, the accuracy is almost constant and is around $15 \, \%$. For $N_{\mathrm{\upmu}} \approx 10^{5.3}$ the reconstruction accuracy is about $24 \, \%$ quickly improving towards higher $N_{\mathrm{\upmu}}$ to below $10 \, \%$ in the relevant energy range.

\section{Analysis Method}
\begin{figure}
\includegraphics[width=\columnwidth]{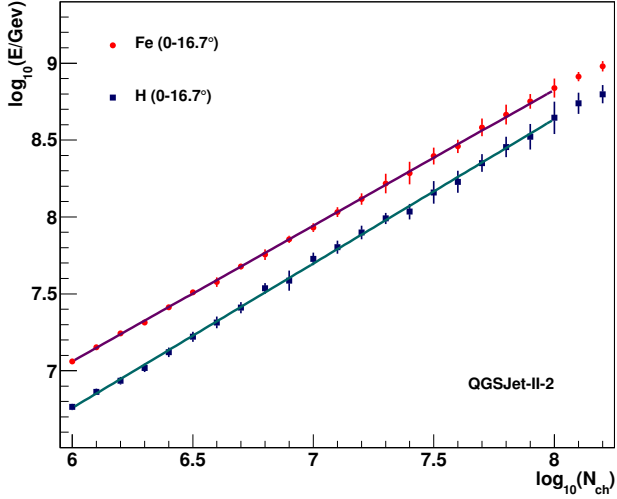}
\caption{The simulated primary energy as a function of the number of charged particles.} 
\label{figEtrueNch}
\end{figure}

\begin{figure}
\centering
\includegraphics[width=\columnwidth]{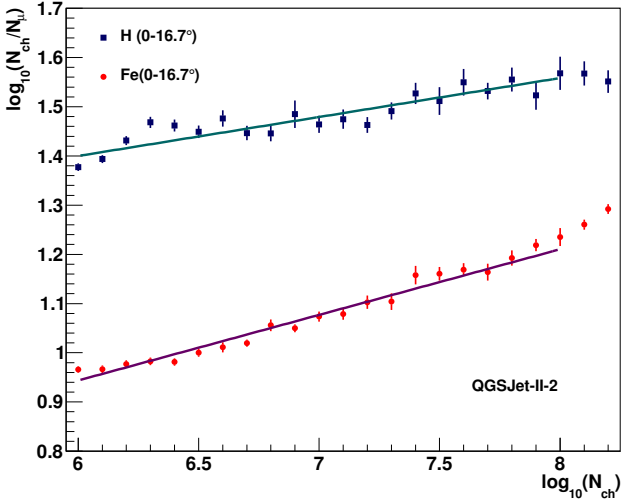}
\caption{The ratio of the reconstructed number of charged particles over the number of muons as a function of the number of charged particles.} 
\label{figNchNmuNch}
\end{figure}

The primary energy is estimated using a combination of $N_{\mathrm{ch}}$ and $N_{\mathrm{\upmu}}$ (Eqs.(\ref{form_energy},~\ref{form_k})). The mass dependence of the reconstructed number of charged particles for a given primary energy is shown in Fig.~\ref{figEtrueNch}, where $N_{\mathrm{ch}}$ is smaller for iron primaries compared to protons. This is mainly due to the lower atmospheric depth at which the first interaction occurs for heavier cosmic rays. This mass dependence of the reconstructed energy is taken into account in terms of the parameter \textit{k} (Eq.(\ref{form_k})) assuming that the two relevant extreme cases are proton and iron primaries. The \textit{k} parameter utilizes the differences in the ratio of $N_{\mathrm{ch}}$ to $N_{\mathrm{\upmu}}$ between proton and iron primaries. The ratios are shown in Fig.~\ref{figNchNmuNch} for the first out of five zenith angular intervals. The separation of the analysis in five zenith angle intervals of equal exposure is done to take the shower attenuation into account. The upper limits are $16.7^\circ$, $24^\circ$, $29.9^\circ$, $35.1^\circ$ and $40^\circ$. 

As a function of $\log_{10}N_{\mathrm{ch}}$, \textit{k} is centered around 0 for protons increasing with the mass of the primary particle to become 1 for iron, see also reference~\cite{Bertaina2012217}.
\begin{figure}
\includegraphics[width=\columnwidth]{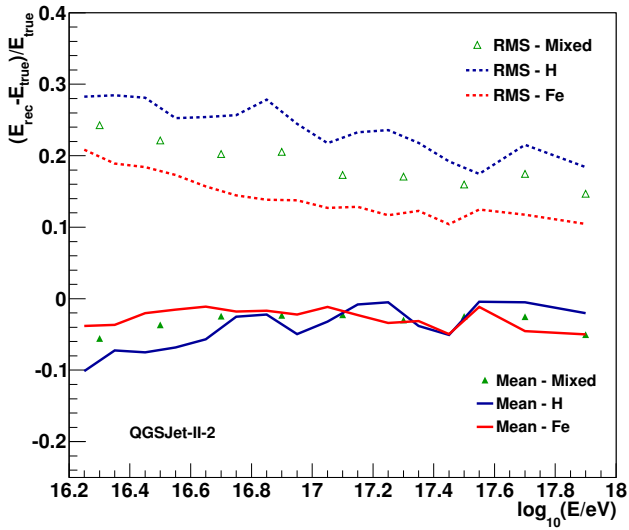}
\caption{The mean difference between the reconstructed energy and the simulated energy and its root mean square.} 
\label{figEAccur}
\end{figure}
\begin{multline}
\log_{10}(E/\mathrm{GeV}) = (a_{\mathrm{H}} + (a_{\mathrm{Fe}} - a_{\mathrm{H}}) \cdot k) \cdot \log_{10}(N_{\mathrm{ch}})\\ + b_{\mathrm{H}} + (b_{\mathrm{Fe}} - b_{\mathrm{H}}) \cdot k
\label{form_energy}
\end{multline}
\begin{equation}
k = \frac{\log_{10}(N_{\mathrm{ch}}/N_{\mathrm{\upmu}})-\log_{10}(N_{\mathrm{ch}}/N_{\mathrm{\upmu}})_{\mathrm{H}}}{\log_{10}(N_{\mathrm{ch}}/N_{\mathrm{\upmu}})_{\mathrm{Fe}}-\log_{10}(N_{\mathrm{ch}}/N_{\mathrm{\upmu}})_{\mathrm{H}}},
\label{form_k}
\end{equation}
\begin{equation}
\log_{10}(N_{\mathrm{ch}}/N_{\mathrm{\upmu}})_{\mathrm{H,Fe}} = c_{\mathrm{H,Fe}} \cdot \log_{10}(N_{\mathrm{ch}}) + d_{\mathrm{H,Fe}},
\end{equation} 
$a$, $b$, $c$ and $d$ are obtained by fitting linear functions to the mean $\log_{10}E_{\mathrm{true}}$ (coefficients $a$ and $b$) and to the mean $\log_{10}(N_{\mathrm{ch}} / N_{\mathrm{\upmu}})$ (coefficients $c$ and $d$) of simulated events as a function of the logarithm of their reconstructed number of charged particles. This is also shown in Fig.~\ref{figEtrueNch} and Fig.~\ref{figNchNmuNch}.
The energy resolution and the mean difference between the reconstructed energy and the simulated energy are shown in Fig.~\ref{figEAccur} for a pure proton/iron component and for a mixed component of five elements (H, He, C, Si, and Fe, $20 \, \% $ each). The negative mean differences ensure that the events are not systematically shifted towards higher energies because of the steep spectrum. At an energy of $10^{16.6} \, \mathrm{eV}$ the energy resolution is about 22.1$ \, \%$ and around 16.1$ \, \%$ at an energy of $10^{17.8} \, \mathrm{eV}$.

\begin{figure}
\includegraphics[width=\columnwidth]{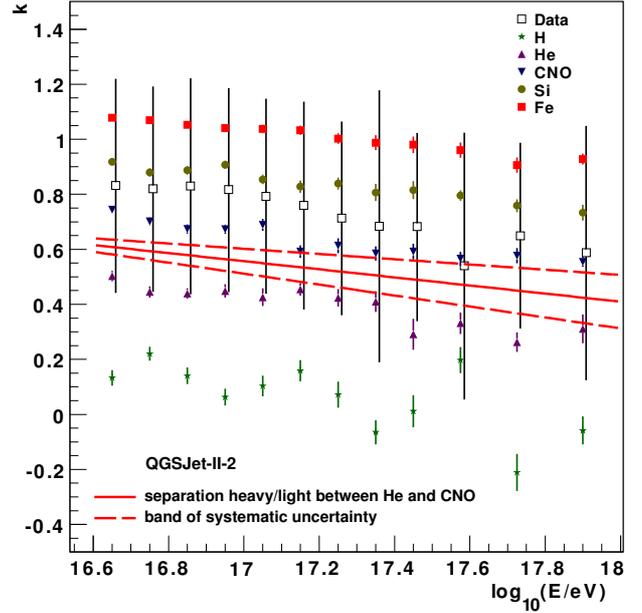}
\caption{The mean values of \textit{k} over the reconstructed energy are shown for events with zenith angles between $0^{\circ}-24^{\circ}$, and five different primaries.
For comparison, the measured data (empty squares) is also shown (shifted from the bin center to the right for better visibility of the error bars). The error bars represent the RMS for the measured data and the error of the mean for the simulated data. The continuous line is used to separate the events into a light and a heavy component, where the dashed lines depict the uncertainty of the separation, taking into account also the reconstruction uncertainty of \textit{k}.} 
\label{figKvsE}
\end{figure}
The mass sensitive parameter \textit{k} can also be used to separate the events into two mass groups. In Fig.~\ref{figKvsE} the mean \textit{k}-value is shown as a function of the reconstructed energy. By fitting a linear function to the $k_{\mathrm{sep}}(E)=[k_{\mathrm{He}}(E)/2 + k_{\mathrm{CNO}}(E)/2]$ distribution the event can be assigned to a mass group by comparing the \textit{k}-value of the event with the corresponding value of the fitted separation line. The dashed lines are used to estimate a possible error of the separation, taking also the reconstruction uncertainty of \textit{k} into account. In order to obtain these lines, $k_{\mathrm{sep}}$ is shifted up/down by the statistical and systematic uncertainties of \textit{k}, before the fit is performed.

\section{Results and Conclusion}
\begin{figure}
\includegraphics[width=\columnwidth]{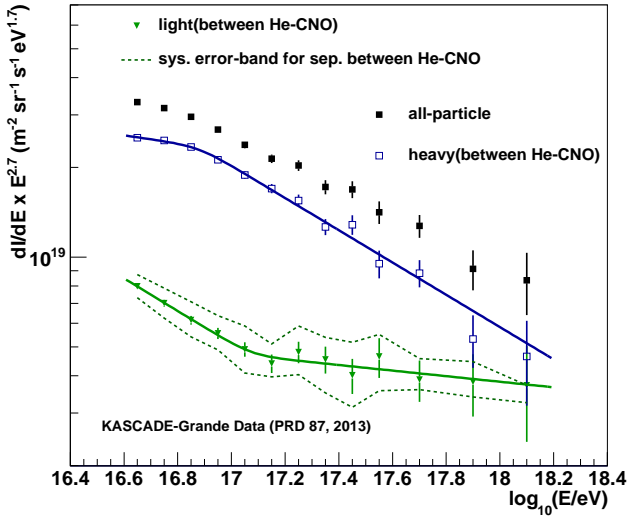}
\caption{The all-particle spectrum and the spectra of heavy and light primaries using the separation line shown in Fig.~\ref{figKvsE}. For the spectrum of light elements, an estimate of a possible error of the separation is also shown.} 
\label{figEspec}
\end{figure}

Applying the analysis to measured data results in the spectra shown in Fig.~\ref{figEspec}. The total systematic uncertainty for the all-particle spectrum is about 11.6$ \, \%$ at an energy of $10^{16.6} \, \mathrm{eV}$ and around 24.4$ \, \%$ at an energy of $10^{17.8} \, \mathrm{eV}$. A detailed description of the considered sources of systematic uncertainties is given in~\cite{Apel2012183}.
For the fits, Eq.(\ref{form_fitfunc})~\cite{TerAntonyan:2000hh} is used.
\begin{equation}
\begin{split}
&\frac{dI}{dE}(E) = I_{0} \cdot E^{\gamma_{1}} \cdot [1 + (\frac{E}{E_{\mathrm{b}}})^{\epsilon}]^{(\gamma_{1} - \gamma_{2})/\epsilon},\\
&\text{$I_{0}$ : normalization factor},\\
&\text{$\gamma_{1/2}$ : index before/after the bending},\\
&\text{$E_{\mathrm{b}}$ : energy of the break position},\\
&\text{$\epsilon$ : smoothness of the break.}\\
\end{split}
\label{form_fitfunc}
\end{equation}
The spectrum of the heavy component exhibits a change of index at $E = 10^{16.88 \pm 0.03} \, \mathrm{eV}$ confirming the \emph{iron-knee} as published in~\cite{prl107}. An ankle-like feature is visible in the spectrum of light elements at an energy of $10^{17.08 \pm 0.08} \, \mathrm{eV}$. At this energy, the spectral index changes from $\gamma_{1} = -3.25 \pm 0.05$ to $\gamma_{2} = -2.79 \pm 0.08$.

It is worth pointing out that the changes in the spectrum of heavy primaries and in the spectrum of light elements are not connected by a bias in the separation or reconstruction procedures. If this would be the case, the bending in the spectrum of heavy particles would be visible at an energy higher than the ankle-like feature in the spectrum of light primaries, because a heavy primary reconstructed as a light particle would be reconstructed with a lower energy. 

The statistical significance that the shown spectrum of light elements cannot be described by a single power law is about $5.8 \, \sigma$. This value corresponds to the Poisson probability $P(N\geq N_{\mathrm{meas}}) = \sum_{k = N_{\mathrm{meas}}}^{\infty}(\frac{N_{\mathrm{exp}}^{k}}{k!} e^{(-N_{\mathrm{exp}})}) \approx 7.23 \times 10^{-09}$ to measure at least $N_{\mathrm{meas}} = 595$ events above the ankle-like feature if $N_{\mathrm{exp}} = 467$ events are expected according to a single power law obtained by a fit to the data points below the bending.

Using EPOS or SIBYLL as underlying hadronic interaction model same results are obtained, but with a variation of the position of the ankle-like structure of the spectrum of light primaries (See e.g.~\cite{MarioICRC2013}).
Due to differences in the high energy hadronic interaction models regarding the number of produced muons, it is also not possible to tell if the shown spectrum of light particles consists mainly of protons and helium primaries or if it is an almost pure proton spectrum. This is shown in~\cite{Apel2013} where a pure proton spectrum simulated using the EPOS (version 1.99~\cite{EPOS}) high energy hadronic interaction model is very similar to a reconstructed spectrum of light primaries using the same EPOS generated events and a QGSJet-II-2 calibration. For simulations using the QGSJet-II-2 model, the reconstructed spectrum of light elements well reproduces a combined proton and helium spectrum.\\

In summary, after separating the events into a light and a heavy component, an ankle-like feature is observed in the spectrum of the light component at an energy of $10^{17.08 \pm 0.08} \, \mathrm{eV}$. The slope index of the underlying power law changes at this energy from $-3.25 \pm 0.05$ to $-2.79 \pm 0.08$, which might be an indication that the transition from galactic to extragalactic origin of cosmic rays starts already in this energy range. 

\section{Acknowledgments}
The authors would like to thank the members of the engineering and technical staff of the KASCADE-Grande collaboration, who contributes to the success of the experiment. The KASCADE-Grande experiment was supported by the BMBF of Germany, the MIUR and INAF of Italy, the Polish Ministry of Science and Higher Education, and the Romanian Authority for Scientific Research
UEFISCDI (grants PNII-IDEI 271/2011 and RU-PD 17/2011). J.C.A.V acknowledges the partial support of CONACyT and the DAAD-Proalmex program (2009-2012).
The present study is supported by the 'Helmholtz Alliance for Astroparticle Physics - HAP' funded by the Initiative and Networking Fund of the Helmholtz Association, Germany.

\clearpage

\end{document}